\documentclass{article}

\usepackage[utf8]{inputenc}
\DeclareUnicodeCharacter{00A0}{~}  
\usepackage{fancyhdr}
\usepackage[hidelinks]{hyperref}
\usepackage{url}
\usepackage{pgfplots,filecontents}
\usepackage{amsmath,amssymb}
\usepackage{caption, booktabs}
\usepackage{physics}
\usepackage[separate-uncertainty=true, range-phrase=-, range-units=single]{siunitx}
\usepackage[left=1in, right=1in, top=1in, bottom=1in]{geometry} 
\usepackage{float}
\usepackage{multirow}
\usepackage[version=4]{mhchem}
\usepackage{graphicx}
\usepackage{gensymb}
\usepackage[margin=1cm, labelfont=bf]{caption}
\usepackage{subfig}
\usepackage[nottoc,numbib]{tocbibind}
\usepackage{wasysym}
\usepackage[english]{babel}
\usepackage[autostyle, english = american]{csquotes}
\MakeOuterQuote{"}
\usepackage{verbatim}
\usepackage{array}
\usepackage{grffile}
\usepackage{accents}
\usepackage{xcolor}
\newcommand*{\thickdot}[1]{%
  \accentset{\mbox{\Large\bfseries .}}{#1}} 
\usepackage{multicol}
\usepackage{authblk}
\usepackage[style=ieee, citestyle=numeric-comp, backend=biber]{biblatex}
\begin{document}

\title{Modeling interspur interactions as a potential mechanism of the FLASH effect.}

\author[1,2,3]{Alexander Baikalov}
\author[3]{Ramin Abolfath}
\author[3]{Radhe Mohan}
\author[3,4]{Emil Schüler}
\author[1]{Jan J. Wilkens}
\author[1,2]{Stefan Bartzsch}

\affil[1]{Technical University of Munich, School of Medicine and Klinikum rechts der Isar, Department of Radiation Oncology, Munich, Germany}
\affil[2]{Helmholtz Zentrum München GmbH, German Research Center for Environmental Health, Institute of Radiation Medicine, Neuherberg, Germany}
\affil[3]{University of Texas MD Anderson Cancer Center, Department of Radiation Physics, Houston, Texas, United States of America}
\affil[4]{Graduate School of Biomedical Sciences, University of Texas, Houston, Texas, United States of America}

\date{\today}

\maketitle

\section*{Abstract}
\label{Abstract}
The mechanism responsible for the FLASH effect, normal tissue sparing by ultra-high dose rate (UHDR) irradiation with isoeffective tumor control compared to conventional dose rate (CDR) irradiation, remains undetermined. Here we investigate the contribution of interspur interactions (interactions between radiolytic species of individual particle tracks) to overall radiochemical interactions as a function of irradiation parameters, and suggest an increase in interspur interaction as a potential mechanism for tissue sparing in FLASH radiation therapy. We construct a model that analytically represents the spatiotemporal distribution of spurs in a target volume as a function of irradiation parameters (e.g. dose, dose rate, linear energy transfer), and quantifies the effect of interspur interactions on the ongoing radiochemistry. Spurs evolve under a simplified reaction-diffusion equation with parameters based on Monte Carlo simulations, and interspur interaction is quantified by calculating the expected values of interspur overlap in the target.

The model demonstrates that for any set of irradiation parameters, a minimum critical dose and dose rate are necessary to induce significant interspur interaction, and that interspur interactions correlate negatively with beam linear energy transfer at a fixed dose. The model suggests optimal beam parameters, including dose, dose rate, linear energy transfer, and pulse structure, to maximize interspur interactions. Depending on the rate of radical scavenging in the target, which limits interspur interaction, this model predicts that the irradiation parameters necessary to elicit the FLASH effect may coincide with an onset of significant interspur interactions, suggesting that interspur interaction may be the underlying mechanism of the FLASH effect.

\section{Introduction}
\label{Introduction}

The past decade has witnessed a surge in investigation into the effects of high dose rate irradiation on tissue toxicity. Many studies, both \textit{in vitro} and \textit{in vivo}, and recently in a human patient, have reported normal tissue sparing by irradiation at ultra-high dose rates (UHDRs, $\gtrapprox \SI{40}{Gy/s}$) relative to irradiation at conventional dose rates (CDRs, $\approx \SI{0.01}{Gy/s}$) while maintaining equivalent tumor control probability \cite{Favaudon_2014, Bourhis_2019_Clinical, Vozenin_2019_Cats, Levy_2020, Montay-Gruel_2018, Zlobinskaya_2014}. This biological effect, known as the FLASH effect, presents significant benefits to clinical radiotherapy applications. However, clinical application of the FLASH effect is currently hindered by an incomplete understanding of both the precise irradiation parameters necessary to elicit a reproducible FLASH effect as well as the underlying mechanisms \cite{Wu_2021, Schuler_2022, Bourhis_2019_Clinical}.

Analyses of studies reporting a FLASH effect or lack thereof have provided some approximate constraints on the irradiation parameters required to induce an observable FLASH effect: an average dose rate $\gtrapprox \SI{40}{Gy/s}$, a total dose $\gtrapprox \SI{10}{Gy}$, and thus an exposure time of $\lessapprox \SI{0.2}{s}$~\cite{Wilson_2020, Vozenin_2019_SleepingBeauty}. These constraints are by no means absolute nor do they guarantee an observable FLASH effect, as evidenced by multiple studies unable to observe a FLASH effect~\cite{Wilson_2020}. In fact, many other parameters, such as pulse width, pulse frequency, tissue type and oxygenation, radiation modality, field size, etc., may play a role. Notwithstanding, the FLASH effect has been reported in many different tissue types including mouse brain~\cite{Montay-Gruel_2019, Montay-Gruel_2018, Montay-Gruel_2017, Simmons_2019}, zebrafish embryos~\cite{Montay-Gruel_2019}, mini-pig skin~\cite{Vozenin_2019_Cats}, cat skin~\cite{Vozenin_2019_Cats}, and human skin~\cite{Bourhis_2019_FirstPatient}, and for different radiation modalities such as electrons~\cite{Montay-Gruel_2017, Favaudon_2014, Vozenin_2019_Cats}, x-rays~\cite{Montay-Gruel_2018}, and protons~\cite{Abel_2019, Buonanno_2019, Rama_2019}.

\begin{figure}[H]%
    \centering
   {\includegraphics[width=12cm]{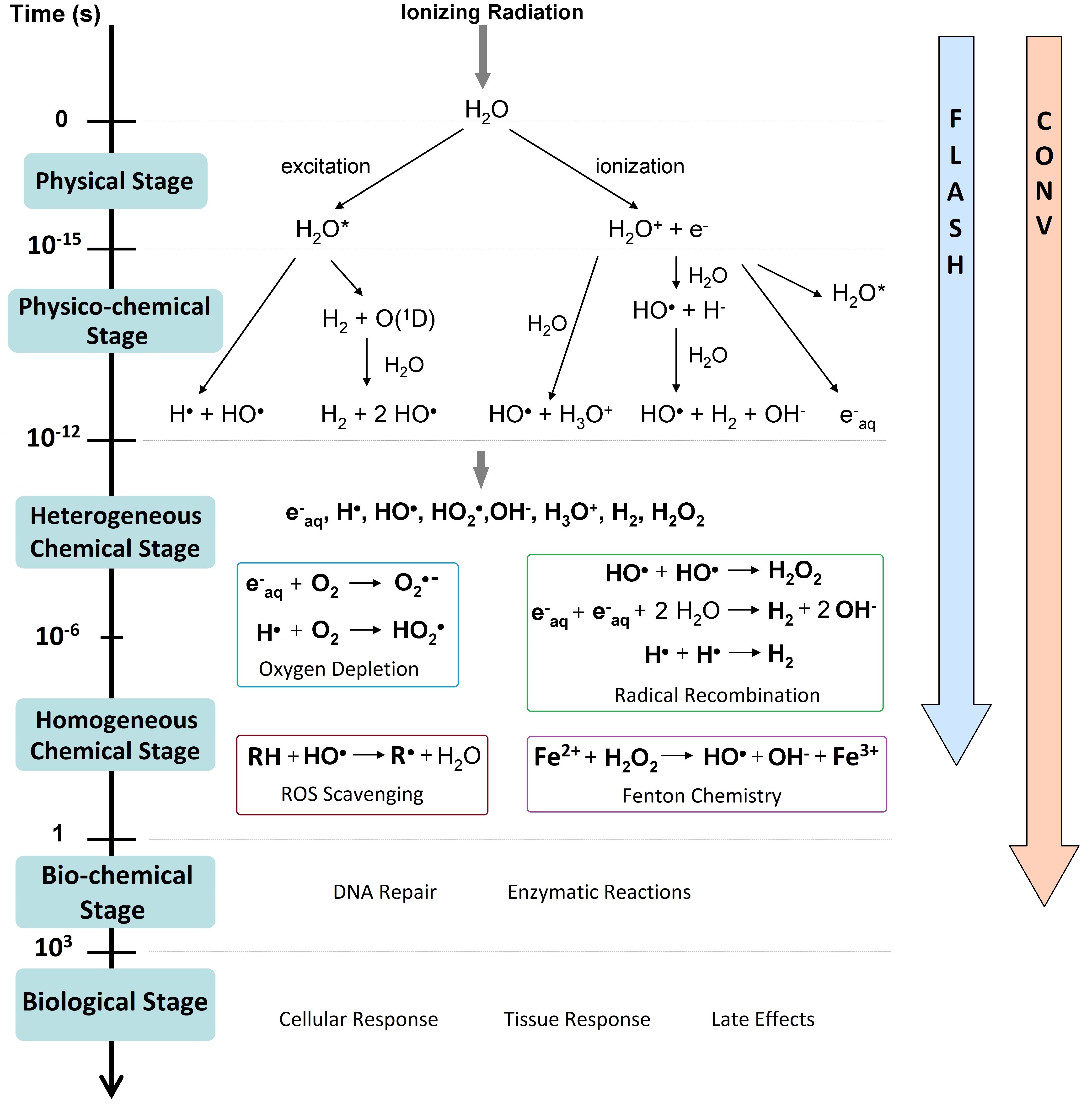}}%
    \caption{Relevant processes of the different approximate time scales of radiation effects in tissue compared with the standard exposure times for FLASH and conventional (CONV) irradiation. The chemical stage can be divided into the heterogeneous and homogeneous chemical stages, as is shown in this figure; however, the listed processes that occur during these phases are not strictly divided and can occur throughout the entire chemical phase. Figure adapted from Le Caër~\cite{LeCaer_2011}.}%
    \label{Radiochemistry Time Scales Figure}%
\end{figure}

Unraveling the mechanism responsible for the FLASH effect (FLASH mechanism) would greatly accelerate the clinical translation of FLASH radiotherapy (RT), e.g. by defining optimum irradiation parameters. Comparing the time scale of radiation effects in tissue (see Figure \ref{Radiochemistry Time Scales Figure}) to irradiation parameters that elicited a FLASH effect in experiments, radiochemical interactions seem to be the most likely source of the FLASH effect. For instance, Montay-Gruel et al.~\ showed an induction of the FLASH effect \textit{in vivo} when a $\SI{10}{Gy}$ exposure was shortened from $\SI{0.5}{s}$ to approximately $\SI{0.2}{s}$~\cite{Montay-Gruel_2017}. Biological responses, although certainly subsequently affected by FLASH irradiation, are unlikely to be the source of the FLASH effect as they occur at time scales too large (e.g. $\approx\SI{e3}{s}$ for DNA repair~\cite{Joubert_2006}) to account for temporal irradiation changes on the order of $\SIrange{0.5}{0.2}{s}$. On the other hand, considering that the physical and physico-chemical stages of a single track are finished within a picosecond and take place on length scales of a few nanometers, interactions between separate tracks at these extremely early stages are negligible. For comparison, only approximately one particle track arrived per $\SI{3.2}{\mu m^2}$ and per $\SI{26}{ns}$ on the target on average in the aforementioned FLASH study for a dose of $\SI{10}{Gy}$ and an exposure time of $\SI{0.2}{s}$. These quantities were derived assuming a cubical water target of $\SI{5}{mm}$ side length to emulate a mouse brain, the target used in the study, and an average energy deposition in that target per primary electron of $\SI{1}{MeV}$, which was calculated via electron stopping power data~\cite{NIST}.

The most prominent hypotheses of FLASH mechanisms in the literature do indeed point to radiochemical effects caused by the shortened time scale of production of water radiolysis products. An overview of the relevant processes of the radiochemical phase is given in Figure \ref{Radiochemistry Time Scales Figure}. For example, the oxygen depletion hypothesis suggests that radiolytic oxygen depletion at UHDR occurs at a rate significantly quicker than tissue oxygen replenishment via diffusion can take place \cite{Montay-Gruel_2019, Adrian_2020, Petersson_2020, Pratx_2019, Durante_2018, Abolfath_2020, Ling_1975, Zakaria_2020, Spitz_2019_Paper}. Therefore, in contrast to CDR, a transient hypoxia of the irradiated tissue may be achieved, rendering it less radiosensitive due to the well-documented hypoxia-induced radioresistance of cellular tissue~\cite{Hall_2006}. Although oxygen concentrations have been shown to significantly affect the onset of the FLASH effect~\cite{Adrian_2020, Montay-Gruel_2019}, many studies indicate that oxygen depletion alone cannot explain the FLASH effect as proposed \cite{Lai_2021, Boscolo_2021, Labarbe_2020, Jansen_2021}.

A variety of different FLASH mechanisms involving altered interaction phenomena among water radiolysis products at UHDR have been proposed. For example, increased rates of radical recombination interactions (e.g. $\ce{^.OH + ^.OH -> H2O2}$) could decrease the yields of radicals available for biological damage \cite{Abolfath_2020, Jansen_2021, Koch_2019}. Additionally, higher radical concentrations directly after UHDR irradiation may maximize the differences between normal and tumor tissue metabolism and radical removal efficiency \cite{Spitz_2019_Paper}.

\begin{figure}[H]%
    \centering
   {\includegraphics[width=10cm]{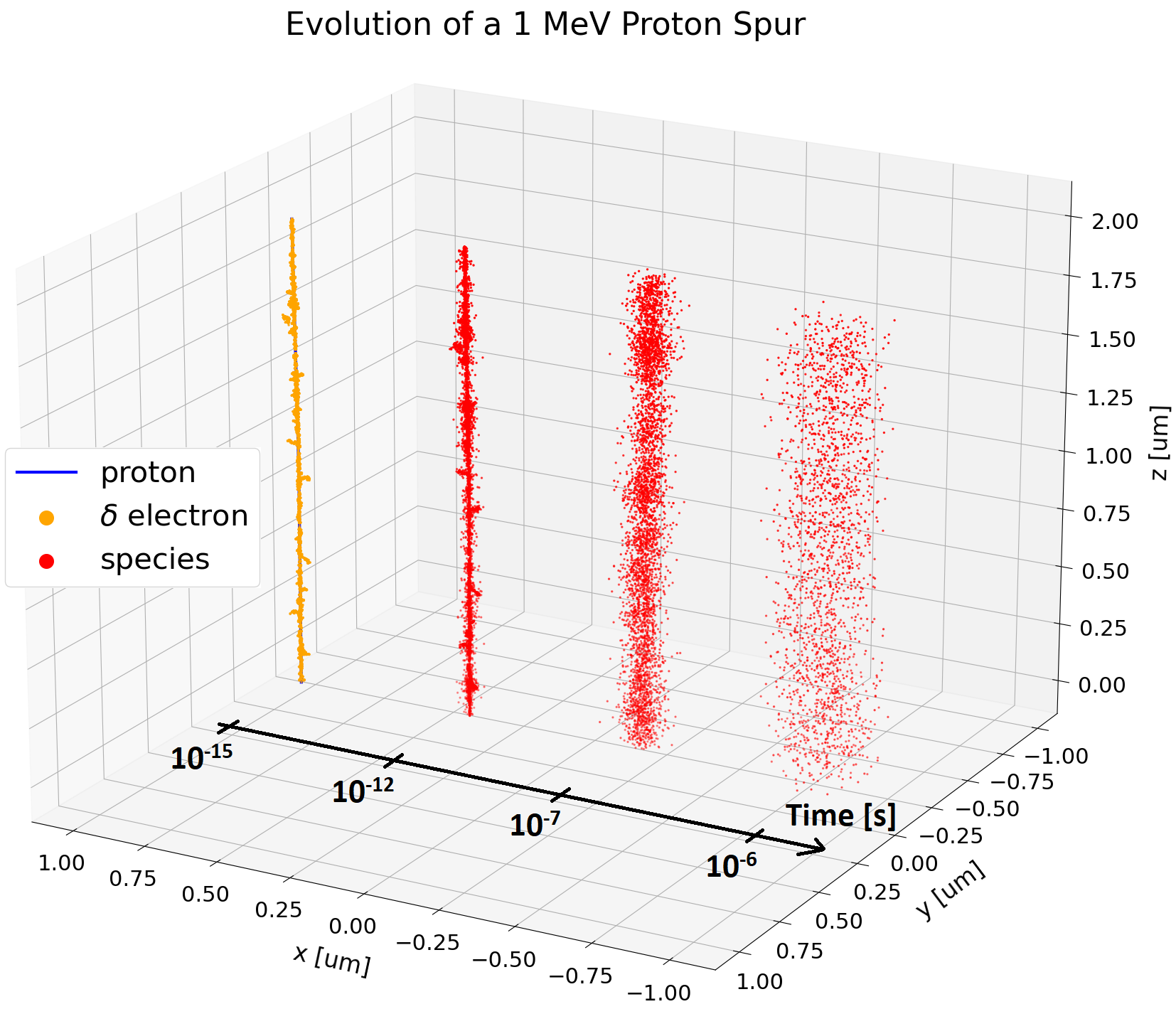}}%
    \caption{Spatial distribution of a $\SI{1}{MeV}$ proton track in pure water (after $\SI{e-15}{s}$) and its resulting spur at different time points ($\SI{e-12}{s}$ - $\SI{e-6}{s}$). The spur is shifted spatially along the x-axis for each time point for better visualization. In this figure, the proton track (blue) is hidden by the large number of $\delta$ electrons (orange) it produces, which are primarily responsible for the production of radiochemical species (red). The beam axis is along the positive z axis. Simulations were performed with TOPAS n-Bio \cite{Schuemann_2019}.}%
    \label{Spur Evolution Figure}%
\end{figure}

\newpage

\section{Methods}
\label{Methods}

\subsection{Motivation}
In this work, we examine radiochemical changes at UHDR that could cause the FLASH effect at the transition from the heterogeneous to the homogeneous chemical stage, as described in Figure \ref{Radiochemistry Time Scales Figure}. While radiochemical species are initially highly localized around the particle track (heterogeneous chemical stage), thermal diffusion dissolves this structure leading to a more homogeneous spread of these radiolysis products. We refer to the collection of radiochemical species produced by a single particle track as that particle's spur. The interactions that these species undergo can be classified into intraspur, spur-environment, and interspur interactions. Every spur experiences intraspur interactions (i.e. interactions among the radiochemical species of a single spur) while undergoing an outward diffusion due to the thermal motion of its species, as depicted in Figure \ref{Spur Evolution Figure}. Intraspur interactions result in a net concentration decay of the species the spur comprises, most notably of the hydroxyl radical ($\ce{^.OH}$), which is responsible for the majority of radiation-induced DNA damage \cite{Roots_1975}. Each spur also experiences interactions with its local aqueous environment, e.g. with dissolved oxygen or cellular scavengers. Lastly, if two or more spurs are located close enough in space and time, they can undergo interspur interactions.

When considering the aforementioned UHDR to FLASH irradiation transition ($\SIrange{0.5}{0.2}{s}$) from the perspective of the chemical stage of spurs, the average spatial distribution of spurs does not change. Instead, only their temporal distribution is changed, as spurs arrive in quicker succession. While intraspur interactions remain unaffected by a change in dose rate, spur-environment and interspur interactions may change with dose rate. The oxygen depletion hypothesis is an example of a spur-environment interaction affected by dose rate. Increasing the temporal density of spurs increases the probability of interspur interactions. Hypotheses for FLASH mechanisms involving increased radical interactions due to shortened exposure times (increased dose rates at a fixed dose) must therefore attribute the increase in radical interactions to either an indirect effect of altered spur-environment interactions, or to increased interspur interactions. In this context, interspur interactions also refer to interspur interactions that occur during the homogeneous chemistry phase, long after the original track structures have dissolved.

Here, we present an analytical model to quantify intraspur and interspur interactions as a function of irradiation parameters. A discussion of the validities and consequences of the assumptions made in this model is given in Section \ref{Discussion}.

\subsection{Spur Definition}
\label{Methods/Definition}
In this model, spurs are assumed to be perfectly straight and homogeneous along the beam axis, and can thus be described in the 2-dimensional plane orthogonal to the beam axis with the position vector $\vec{r}=(x,y)$. An analogous 3-dimensional model, more representative of sparsely ionizing radiation, is discussed in Section \ref{Discussion/Sparsely}.

We consider a system of $N_s$ spurs distributed across a target. An arbitrary spur at position $\vec{r_i}$ and creation time $t_i$ is defined by a probability density function (PDF) $c_i(\vec{r}, t)$ describing the spatio-temporal distribution of the radiochemical species that make up the spur. The spatio-temporal evolution of the spur is governed by a reaction-diffusion equation comprising a cylindrically-symmetric diffusion term with diffusion coefficient $\alpha$, a decay term with decay constant $k_s$, and an interspur interaction term with reaction rate $k_r$:
\begin{equation}
    \frac{\partial}{\partial t}c_i(\vec{r},t) = \alpha\nabla^2 c_i(\vec{r},t) - k_s c_i(\vec{r},t) - k_r \sum_{i \neq j}^{N_s} c_i(\vec{r},t) c_j(\vec{r},t) \ .
    \label{DGL}
\end{equation}
The decay term represents the effective result of all intraspur and spur-environment interactions; thus, the spur decay constant $k_s$ is a model parameter that fixes the lifetime of a spur depending on the spur's own characteristics and those of its environment. The interspur interaction term considers the interaction with every other spur on the target. In this work we only consider the limit of weakly-interacting spurs and thus neglect this interspur interaction term.

If we set $c_0$ to be the initial total number of species in any spur, the solution to this partial differential equation (see Section \ref{Appendix/PDF}), neglecting interspur interactions, is a normal distribution about the arrival point of the spur that broadens and decays with time:
\begin{equation}
    c_i(\vec{r},t) = c_0 \cdot \frac{e^{-\frac{\abs{\vec{r}-\vec{r_i}}^2}{4\alpha (t-t_i)} - k_s (t-t_i)}}{4\pi \alpha (t-t_i)} \ ,
    \label{PDF}
\end{equation}

One disadvantage of describing the spur by this PDF is that at very short time scales, the normal distribution becomes infinitely dense, which is an unrealistic representation of the spur, and leads to problems when calculating intraspur interactions. In reality, the species of a spur already have a finitely spread distribution at the moment of production, caused by spatial variance in the primary particle's path as well as a spatial spread in the production of chemical species in the physico-chemical phase. In order to address this problem, we apply a temporal shift to give each spur a minimum effective age $\tau_0$, which represents its non-zero spatial distribution upon creation. Thus, the PDF of an arbitrary spur centered at $\vec{r_i}$ for $t>t_i$ is
\begin{equation}
    c_i(\vec{r},t) = c_0 \cdot \frac{e^{-\frac{\abs{\vec{r}-\vec{r_i}}^2}{4\alpha (t-t_i + \tau_0)} - k_s (t-t_i)}}{4\pi \alpha (t-t_i + \tau_0)}\ .
    \label{PDF Arbitrary}
\end{equation} 
Quantitative values for the diffusion coefficient $\alpha$, minimum spur age $\tau_0$, and decay constant $k_s$ can be approximated by fitting characteristics of the analytical spur to the MC simulated spur, as shown in Figures \ref{Figures/Intraspur}a and \ref{Figures/Intraspur}b. 

\subsection{Spur Distribution}
\label{Methods/Distribution}
For simplicity, the beam is assumed to have a constant dose rate over the exposure time $T$, and thus the PDF of the creation time $t_i$ of the $i$th spur at time $t$ is
\begin{equation}
    P_t(t_i) =
    \begin{cases} 
      \frac{1}{\min(t, T)} \quad & \text{for} \quad 0\leq t_i\leq \min(t,T)\\
      0 \quad & \text{elsewhere}
   \end{cases}
   \ .
   \label{P(t_i) Constant Beam Equation}
\end{equation}
The expected total number of spurs on the target over time $N_s(t)$ can be expressed with the total energy deposited in the target $E$ and the average energy deposited per spur $E_s$ for a target of density $\rho$, thickness $z$ (measured along the beam axis), and cross-sectional area $A$: 
\begin{align}
    N_s(t) &= \frac{E}{E_s} \cdot \frac{\min(t, T)}{T}\\
    &= \frac{D \cdot A \cdot \rho \cdot z}{E_s} \cdot \frac{\min(t, T)}{T}\\
    &= \frac{D \cdot A \cdot \rho}{L} \cdot \frac{\min(t, T)}{T}\ .
    \label{N_s(t) Equation}
\end{align}
In the last step we introduce the linear energy transfer (LET) $L=E_s/z$, which will depend on the radiation quality.

Assuming a spatially homogeneous beam across the target area $A$, the expected number of spurs within any circular area of radius $R$ can be expressed as a fraction of the total number of spurs on the target:
\begin{equation}
    N_R(t) = N_s(t) \cdot \frac{\pi R^2}{A}\\
    \label{N_R(t) Equation}
\end{equation}
The PDF of the displacement $s$ between one arbitrary spur and any other spur within a radius $R$ is given by the ratio between the area element $2\pi s$ and the total area $\pi R^2$:
\begin{equation}
    P_s(s) = \frac{2 s}{R^2} \ , \quad s \leq R \ .
    \label{P(s) Equation}
\end{equation}

\subsection{Interaction Quantification}
\label{Methods/Interaction}
The quantification of interspur interaction is a critical component of this model. A measure is needed that takes into account the spatial variations of different spurs, and reflects a physical quantity relevant to the radiochemistry at hand. To this end, the interaction rate $\omega$ is first defined, which yields the rate of change of the quantity of species at time $t$ due to the interaction of two spurs assuming second order reactions with reaction rate $k_r$:
\begin{align}
    \omega_{1,2}(t) &=  \int k_r \cdot c_1(\vec{r},t) \cdot c_2(\vec{r},t) \, dV\\
    & = k_r \cdot c_0^2 \cdot \frac{e^{-\frac{\abs{\vec{r_1}-\vec{r_2}}^2}{4\alpha(2t-t_1-t_2+2\tau_0)} - k_s(2t-t_1-t_2)}}{4\pi \alpha(2t-t_1-t_2+2\tau_0)} \ .
    \label{Interaction Rate Density Equation}
\end{align}

Integrating the interaction rate in Equation~\ref{Interaction Rate Density Equation} over time yields the total interspur conversion $I$, which represents the net change in the quantity of species due to the interspur interaction over the relevant time period:
\begin{equation}
    I_{1,2}(t) = \int_0^t \omega_{1,2}(t') \, dt' \ .
\end{equation}
$I$ will be the measure of interest in this model, as it represents an empirically measurable quantity.

\subsection{Interspur Interactions}
\label{Methods/Interspur}
The expected interaction rate of an arbitrary spur $a$ with all neighboring spurs within an interaction volume of radius $R$ can be found by multiplying the number of spurs inside that interaction volume $N_R$ by the expected interaction rate between one arbitrary spur and another within that volume,
\begin{equation}
     \sum_i \, \omega_{a,i}(t) = N_R(t) \cdot \langle \omega_{1,2}(t) \rangle \ ,
     \label{Summed Average Interaction Rate Density Equation}
\end{equation}

The total interspur conversion of all spurs in the target volume is then
\begin{equation}
    I_\text{inter}(t) = \frac{N_s(t)}{2} \cdot \int_{0}^t \lim_{R\rightarrow\infty} N_R(t') \cdot \langle \omega_{1,2}(t') \rangle  \, dt' \ ,
    \label{Total Inter-spur Interaction Equation}
\end{equation}
where $R$ approaches infinity under the assumption that each spur is effectively within an infinite isotropic volume. The factor $\frac{1}{2}$ ensures that each interspur interaction is not double-counted.

Evaluating this expression (see Section
\ref{Appendix/Interspur}) yields
\begin{equation}
    I_\text{inter}(t) =
    \begin{cases} 
        \lim_{\epsilon\rightarrow 0} \frac{B}{2} \cdot t \cdot \left[\ln(\frac{t}{\epsilon}) - \Gamma(0, 2k_st) + 2\Gamma(0, k_st) + \Gamma(0, 2k_s\epsilon) - 2\Gamma(0, k_s\epsilon)\right] & \text{for} \quad t\leq T\\[10pt]
        I_\text{inter}(T) + \frac{B}{2} \cdot \frac{(e^{k_sT}-1)^2 }{2k_s} \cdot (e^{-2k_s T}-e^{-2k_st}) & \text{for} \quad t> T
    \end{cases} \ ,
    \label{P_inter(t)}
\end{equation}
where
\begin{equation}
    B = \frac{N_s(t)^2}{A} \cdot \frac{k_r}{k_s^2} \cdot c_0^2 \cdot \left(\frac{1}{\min(t, T)}\right)^2 \ ,
\end{equation}
and $\Gamma(s, x)$ is the incomplete upper gamma function given by
\begin{equation}
    \Gamma(s, x) = \int_x^\infty t^{s-1}e^{-t}\, dt \ .
    \label{Gamma_Func}
\end{equation}

\subsection{Intraspur Interactions}
\label{Methods/Instraspur}
The total intraspur conversion across the entire target volume is simply a sum of each spur's self-interaction
\begin{equation}
    I_\text{intra}(t) = \sum_i^{N_s(t)} \int_{0}^{t} \omega_{i,i}(t') \, dt' \ .
    \label{Total Intra-spur Interaction Equation}
\end{equation}
This expression can be simplified with a suitable approximation, removing the need for tedious summation over all spurs. Due to a spur's outward diffusion, the amount of intraspur conversion converges over time to a maximum value. If this convergence time is negligible compared to the total exposure time, the total intraspur conversion can be approximated as a linear increase over the exposure time towards the total maximum amount of intraspur conversion of all spurs on the target:
\begin{align}
    I_\text{intra}(t) &\approx N_s(t) \cdot \int_0^\infty \omega_{i,i}(t) \, dt\\
    &= N_s(t) \cdot \int_0^\infty k_r \cdot c_0^2 \cdot \frac{ e^{-2k_st}}{8\pi \alpha(t + \tau_0)}  \, dt  \\
    &= N_s(t) \cdot k_r\cdot c_0^2 \cdot \frac{e^{2k_s\tau_0}}{8\pi \alpha} \cdot \Gamma\left(0,\,2k_s\tau_0\right) \ .
\end{align}

\subsection{Ratio of Interspur to Intraspur Interactions}
\label{Methods/Ratio}
In order to quantify the extent to which interspur interactions affect the overall radiochemistry, the ratio of interspur to intraspur interaction is used in this model as the main measure of interest:
\begin{equation}
    \Phi(t) = \frac{I_\text{inter}(t)}{I_\text{intra}(t)} \ ,
    \label{Ratio Equation}
\end{equation}
whereby we define significant interspur interaction as $\Phi(t) \geq 1$.

Evaluating this ratio with Equations \ref{Total Inter-spur Interaction Equation}, \ref{Total Intra-spur Interaction Equation}, and \ref{N_s(t) Equation} yields
\begin{equation}
    \Phi(t) = 2\pi \alpha \cdot \frac{D \cdot \rho}{L} \cdot \frac{e^{-2k_s\tau_0}}{k_s^2 \cdot \Gamma(0, 2k_s \tau_0)} \cdot \frac{1}{T} \cdot
    \begin{cases} 
         f_1(t, k_s) & \text{for} \quad t\leq T\\[10pt]
         f_1(T, k_s) + \frac{(e^{k_sT}-1)^2}{k_s T}\left(e^{-2k_s T}-e^{-2k_st}\right)  & \text{for} \quad t> T
    \end{cases}
    \label{Phi(t) Equation}
\end{equation}
where
\begin{equation}
    f_1(t', k_s) = 2 \cdot \lim_{\epsilon\rightarrow 0} \left[\ln(\frac{t'}{\epsilon}) - \Gamma(0, 2k_st') + 2\Gamma(0, k_st') + \Gamma(0, 2k_s\epsilon) - 2\Gamma(0, k_s\epsilon)\right] \ ,
\end{equation}
and $\Gamma(s, x)$ is defined as in Equation \ref{Gamma_Func}. Table \ref{Tables/Parameters} summarizes all variables and their definitions.\\

The primary particle fluence can be expressed using Equation \ref{N_s(t) Equation} as
\begin{align}
    F &= \frac{N_s}{A}\\
    &= \frac{D \cdot \rho}{L} \ .
    \label{Fluence Equation}
\end{align}

The value of $\Phi(t)$ for $t\gg T$ is of interest in order to analyze the final ratio of interspur to intraspur interactions after all spurs have decayed and no more interactions can occur. This value, using Equations \ref{Phi(t) Equation} and \ref{Fluence Equation}, is given by
\begin{align}
    \Phi_{t\gg T} &= \lim_{t\rightarrow\infty} \Phi(t) \\
    &= \alpha \cdot F \cdot f_2(T, k_s, \tau_0) \ ,
    \label{Phi(t>>T) Equation}
\end{align}
where
\begin{equation}
    f_2(T, k_s, \tau_0) = \frac{2\pi}{T} \cdot \frac{e^{-2k_s\tau_0}}{k_s^2 \cdot \Gamma(0, 2k_s \tau_0)} \cdot \left[f_1(T, k_s) +  \frac{(e^{-k_sT}-1)^2}{k_sT}\right] \ .
\end{equation}

\begin{table}[H]
\centering
\begin{tabular}{c|l}

Variable       & Description    \\ \midrule
$\alpha$       & Spur diffusion coefficient \\ 
$\tau_0$       & Spur minimum age \\ 
$k_s$          & Spur decay constant \\ 
$\rho$         & Target density \\ 	
$L$            & LET \\ 	
$D$            & Dose \\
$T$            & Exposure time \\
$F$            & Primary particle fluence\\
\end{tabular}
\caption{Variables described in this model.}
\label{Tables/Parameters}
\end{table}

\subsection{Monte Carlo Simulation of a Particle Spur}
\label{Methods/MC}
\begin{figure}[H]%
    \centering
    \subfloat[][\centering]{\includegraphics[width=8cm]{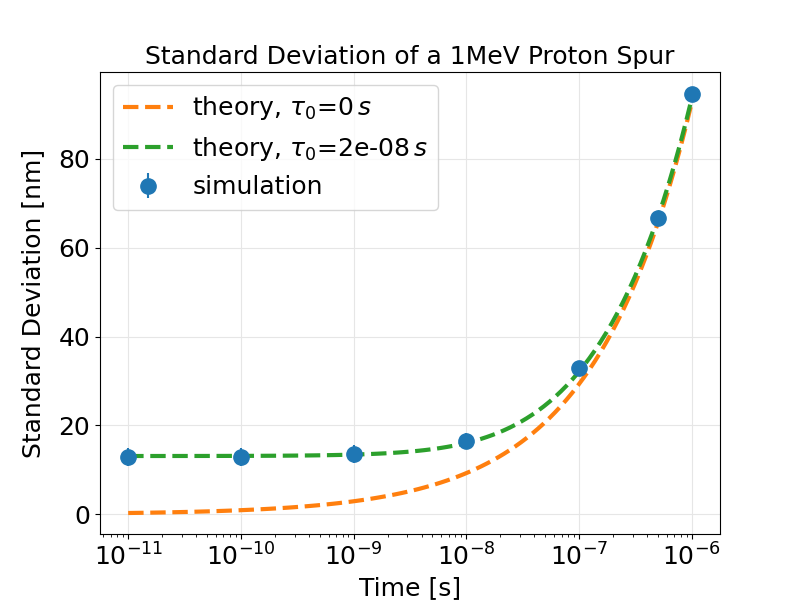}}%
    \quad
    \subfloat[][\centering]{\includegraphics[width=8cm]{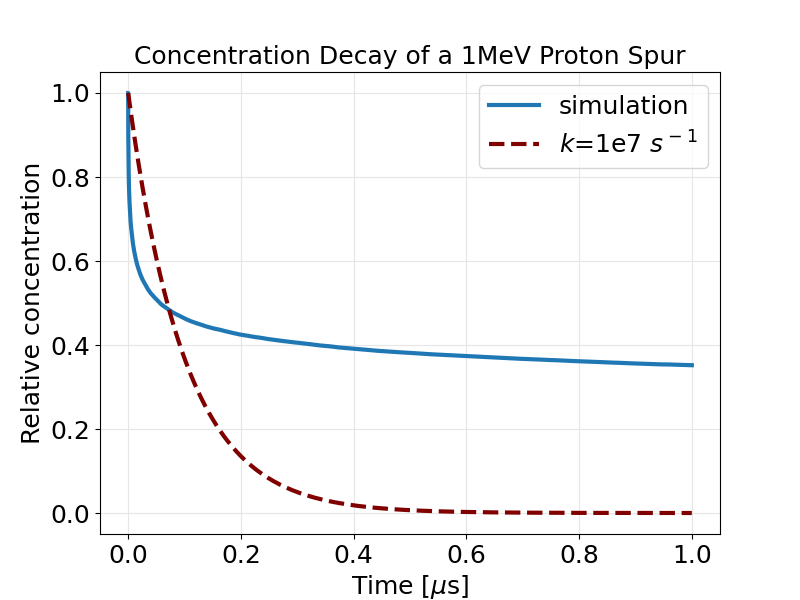}}%
    \caption{Evolution characteristics of the species $\ce{^.OH}$, $\ce{e_{aq}^-}$, $\ce{H^.}$, $\ce{H2O2}$, and $\ce{OH^-}$ from a $\SI{1}{MeV}$ proton spur simulated with TOPAS n-Bio \cite{Schuemann_2019}. \textbf{(a)} Standard deviations of the radial displacements of the spur (blue points) alongside the theoretical standard deviations (dashed lines) assuming a normal distribution ($\sigma = \sqrt{2\alpha (t+\tau_0)\,}$). Including the minimum spur age correction $\tau_0$ results in a better fit of the simulated spur, due to its initial non-zero spatial variance. Fitted values for these parameters to the simulated data are $\alpha=\SI{4.35(4)e-9}{m^2/s}$ and $\tau_0=\SI{1.99(8)e-8}{s}$. \textbf{(b)} Concentration decay of the same simulated $\SI{1}{MeV}$ proton spur relative to its initial concentration due to intraspur interactions (blue line) alongside an exponential decay of the form $e^{-kt}$ for reference.}%
    \label{Figures/Intraspur}%
\end{figure}

Monte Carlo (MC) simulations, with the MC radiolysis toolkit TOPAS n-Bio~\cite{Schuemann_2019}, were performed in order to characterize the evolution dynamics of a spur. A single $\SI{1}{MeV}$ proton spur in a pure water target, as depicted in Figure \ref{Spur Evolution Figure}, was analyzed from the physical stage through the end of the chemical stage ($\SI{1}{\mu s}$). The cubical scoring volume, with a $\SI{2}{\mu m}$ side-length, was embedded in the target volume to ensure diffusive equilibrium within the scoring volume at all time points. The resulting diffusive broadening of the spur and concentration decay of the spur due to intraspur interactions are depicted in Figures \ref{Figures/Intraspur}a and \ref{Figures/Intraspur}b, respectively. Fitting the analytical descriptions (Equation \ref{PDF Arbitrary}) of a spur to this MC-simulated spur allows for approximation of the spur parameters $\alpha$ and $\tau_0$.

\newpage
\section{Results}
\label{Results}

For the following analysis of Equations \ref{Phi(t) Equation} and \ref{Phi(t>>T) Equation}, results are shown graphically for different values of the spur decay constant $k_s$ and LET $L$. A large range of potential values of $k_s$ is used, as its realistic value in water or in tissue is uncertain. For the LET, $L\approx\SI{0.2}{keV/\mu m}$ corresponds to a $\SI{6}{MeV}$ electron beam in a $\SI{5}{mm}$ thick water target, which replicates the beam used in an empirical FLASH study~\cite{Montay-Gruel_2017}, while $L\approx\SI{26}{keV/\mu m}$ corresponds to a $\SI{1}{MeV}$ proton. Values for the diffusion constant $\alpha$ and the minimum spur age $\tau_0$ are taken from the fits described in Figure \ref{Figures/Intraspur}a. The values used for these parameters are discussed further in Section \ref{Discussion/Model}.

\begin{figure}[H]%
    \centering
    \subfloat[][\centering]{\includegraphics[width=7cm]{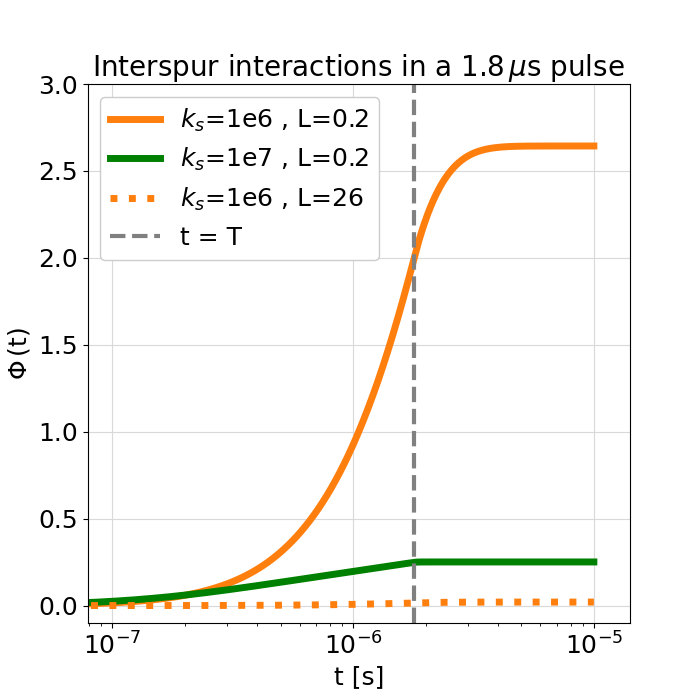}}%
    \quad
    \subfloat[][\centering]{\includegraphics[width=7cm]{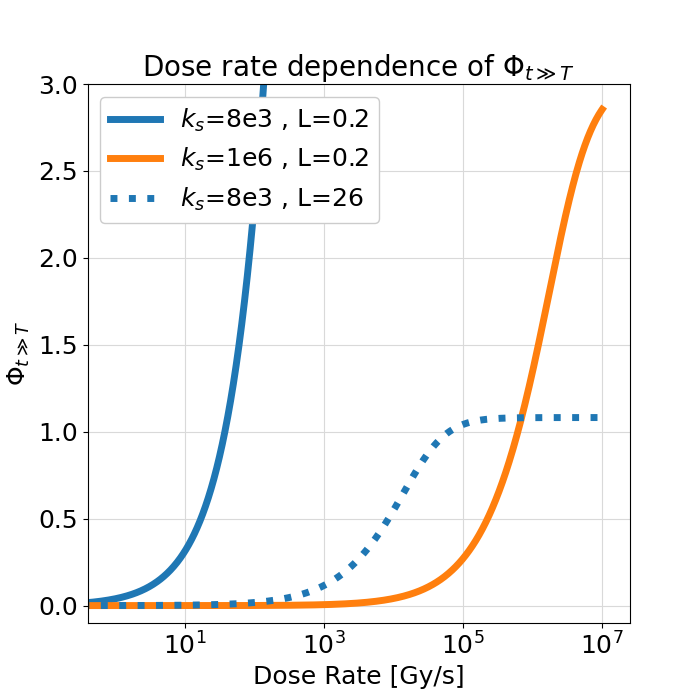}}%
    \caption{\textbf{(a)} The ratio of interspur to intraspur interactions, $\Phi(t)$, during and after a $\SI{1.8}{\mu s}$, $\SI{10}{Gy}$ exposure. \textbf{(b)} The final total ratio of interspur to intraspur interactions, $\Phi_{t\gg T}$, after a $\SI{10}{Gy}$ exposure as a function of dose rate at a fixed dose. Results are shown for different spur decay constants $k_s$ (indicated by color, given in $s^{-1}$) and beam LET (indicated by line style, given in $\SI{}{keV/\mu m}$).}%
    \label{Figures/Electron_Beam_Results}%
\end{figure}

Figure \ref{Figures/Electron_Beam_Results}a depicts the increase in $\Phi(t)$, i.e. the ratio of interspur to intraspur interactions, during and after a single $\SI{1.8}{\mu s}$ pulse exposure of $\SI{10}{Gy}$ for different values of the spur decay constant $k_s$. Initially, $\Phi(t)\approx 0$ as the first few spurs undergo intraspur interactions. As the exposure continues and more spurs are created, begin to broaden, and overlap, interspur interactions become more and more prominent. Once the exposure has ended (indicated by the vertical dashed line), interspur interactions continue to increase until the final remaining spurs have decayed, at which point $\Phi(t)$ approaches its final value, $\Phi_{t\gg T}$. Increasing the spur decay constant $k_s$ reduces interspur interactions more than it reduces intraspur interactions, which can result in interspur interactions never becoming significant, i.e. $\Phi_{t\gg T} < 1$, as seen in Figure \ref{Figures/Electron_Beam_Results}a. Similarly, increasing the beam LET reduces the significance of interspur interactions.

Figure \ref{Figures/Electron_Beam_Results}b depicts the final ratio of interspur to intraspur interactions, $\Phi_{t\gg T}$, across a large range of dose rates at a fixed dose of $\SI{10}{Gy}$. At very low dose rates, where the creation time between spurs is large, interspur interactions are negligible as individual spurs decay before subsequent spurs are created. As the dose rate increases, $\Phi_{t\gg T}$ grows as spurs are more temporally concentrated and interspur interactions become more prominent. If the spur decay constant is sufficiently low, interspur interactions may become significant ($\Phi_{t\gg T}\geq 1$) at some critical dose rate. At very high dose rates, all spurs arrive almost instantaneously relative to their decay lifetimes, and the amount of interspur interaction is determined only by their spatial separation, i.e. the fluence, and the spur decay constant $k_s$. A higher LET results in a decreased fluence, thereby decreasing interspur interaction and increasing the critical dose rate needed to achieve significant interspur interaction. A similar effect occurs when increasing the spur decay constant. In fact, if the spur decay constant or LET are too high, significant interspur interaction may never be possible at any dose rate.

\begin{figure}[H]%
    \centering
    \subfloat[][\centering]{\includegraphics[width=8cm]{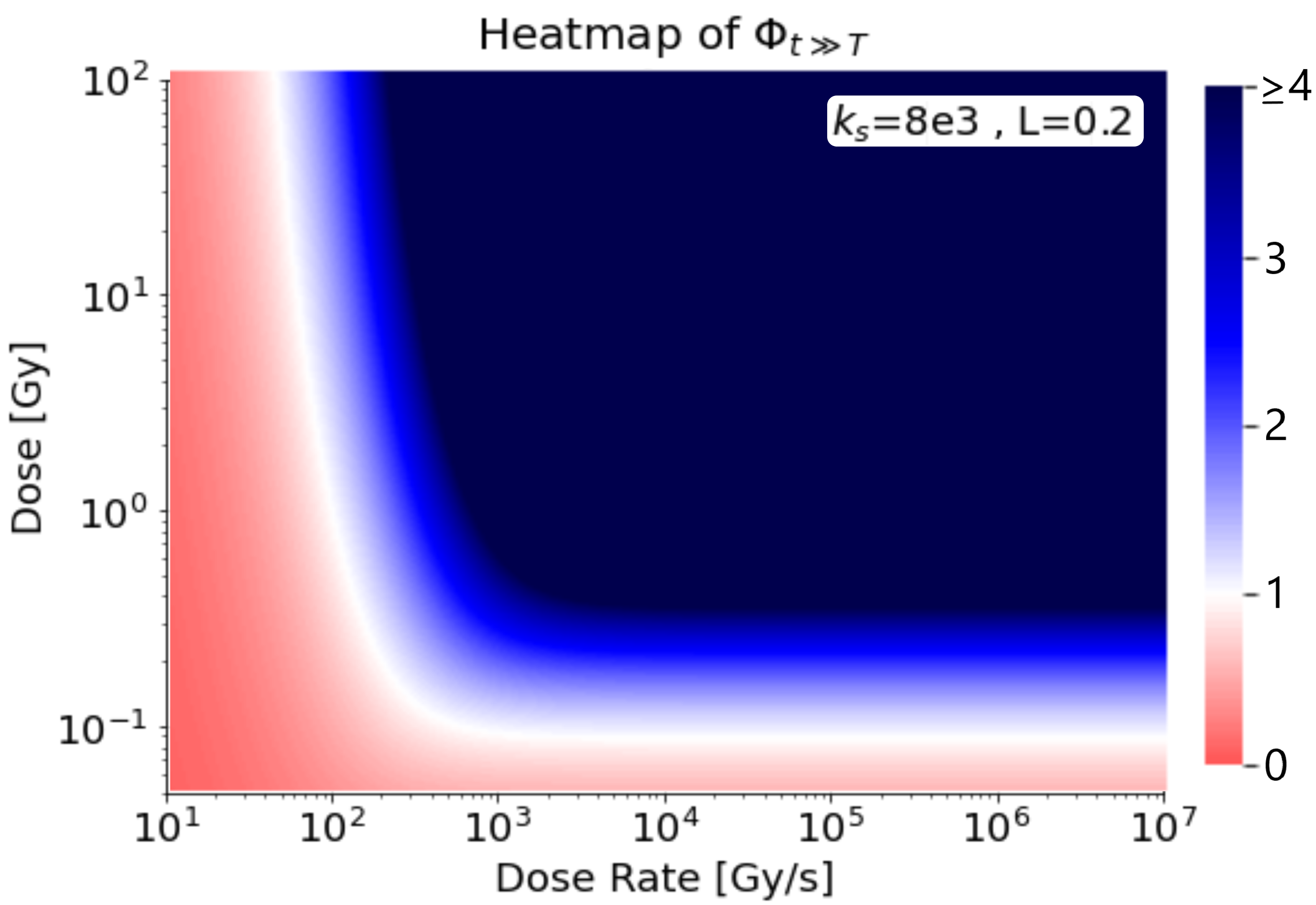}}%
    \quad
    \subfloat[][\centering]{\includegraphics[width=8cm]{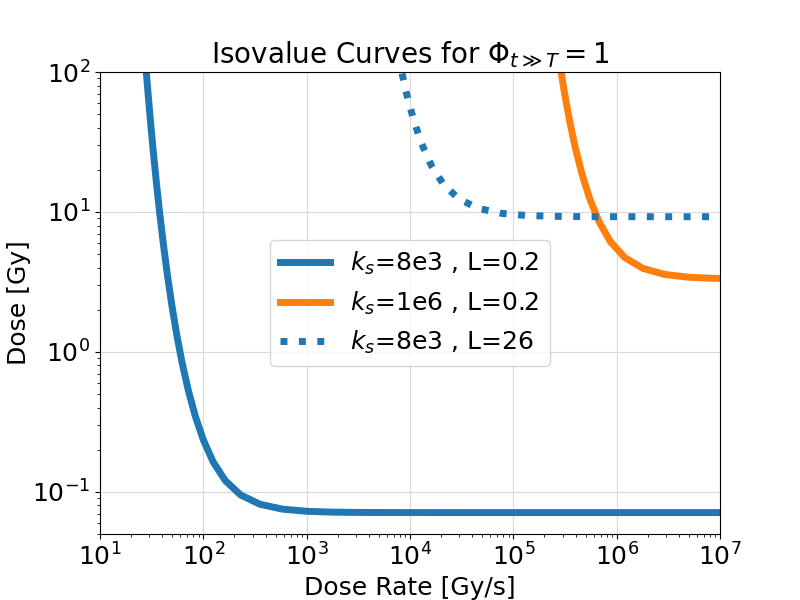}}%
    \caption{\textbf{(a)} Heatmap showing the value of $\Phi_{t\gg T}$ for different dose and dose rate combinations. Interspur interactions are considered significant when $\Phi_{t\gg T} \geq 1$; therefore, the white strip indicates the transition from intraspur-dominated interactions (red) and interspur-dominated interactions (blue). \textbf{(b)} Isovalue curves showing where $\Phi_{t\gg T}=1$ for different spur decay constants and LETs. Results are shown for different spur decay constants $k_s$ (indicated by color, given in $s^{-1}$) and beam LET (indicated by line style, given in $\SI{}{keV/\mu m}$). Increasing the spur decay constant or the LET results in increased doses and dose rates necessary to achieve significant interspur interaction.}%
    \label{Figures/Heatmap_and_Dose_DoseRate}%
\end{figure}

To illustrate the dose and dose rate dependencies of significant interspur interaction simultaneously, Figure \ref{Figures/Heatmap_and_Dose_DoseRate}a depicts the value of $\Phi_{t\gg T}$ for a large range of dose and dose rate combinations. This heatmap demonstrates the transition from intraspur-dominated regions (red) to interspur-dominated regions (blue) of the dose/dose rate parameter space. The isovalue curve at $\Phi_{t\gg T}=1$ (white line) indicates the minimum doses and dose rates necessary to achieve significant interspur interaction. Figure \ref{Figures/Heatmap_and_Dose_DoseRate}b shows this isovalue curve for different values of the spur decay constant $k_s$ and LET $L$. Apparent in this figure is that increasing the spur decay constant or the LET raises the critical doses and dose rates for significant interspur interaction.

\newpage
\section{Discussion}
\label{Discussion}

\subsection{Results}
Equation \ref{Phi(t>>T) Equation} expresses the significance of interspur interactions as a function of irradiation parameters. It demonstrates that the final ratio of interspur to intraspur interactions, $\Phi_{t\gg T}$, is only dependent on two beam parameters: the fluence $F$ and the exposure time $T$, whereby the fluence is determined by the dose $D$ and LET $L$. All other dependencies pertain to parameters specific to the spur: the spur diffusion constant $\alpha$, the minimum spur age $\tau_0$, and the spur decay constant $k_s$.

At fixed values for these spur parameters and a fixed LET, Figure \ref{Figures/Heatmap_and_Dose_DoseRate}a illustrates how interspur interactions depend on the dose and dose rate. For any set of irradiation parameters, there is a minimum dose and a minimum dose rate necessary for significant interspur interactions. These correspond to a minimum fluence and a maximum exposure time, respectively. Thus, interspur interactions are limited in both the spatial (fluence) and the temporal (exposure time) domain. At low dose rates, extremely high doses are necessary to elicit significant interspur interactions. As the dose rate is increased, this critical dose is strongly reduced; a one order of magnitude increase in dose rate may result in a three or more order of magnitude decrease in critical dose. However, once a certain dose rate is reached, interspur interactions becomes dose-rate-independent. Here, the temporal separation between spur arrivals is so short relative to their evolution and interaction, that all spurs arrive effectively instantaneously; thus, interspur interactions are purely dependent on the spurs' spatial separation (determined by the fluence).

Figure \ref{Figures/Heatmap_and_Dose_DoseRate}b depicts the effects of the beam LET and the spur decay constant on $\Phi_{t\gg T}$. A higher-LET beam has a reduced fluence (each particle deposits more energy in the target), and thereby requires a higher minimum dose and dose rate necessary for significant interspur interactions. Applied to proton beams, this model thus predicts that, for the same total dose, the plateau region of a proton beam would induce much stronger interspur interaction than the Bragg peak region. Similarly to an increased LET, an increased spur decay constant $k_s$ results in less interspur interactions due to a quicker decay rate of each spur.

\subsubsection*{Quantitative Considerations}
The descriptions of the results of this model have thus far been purposefully qualitative. Precise quantitative predictions with this model in its current form are not possible due to uncertainties in the values of multiple critical parameters as well as assumptions made in the construction of the model. For example, interspur interactions were arbitrarily set to become significant when $\Phi_{t\gg T}=1$; however, interspur interactions may have significant impacts on the radiochemistry at much lower values of $\Phi_{t\gg T}$. Figure \ref{Figures/Heatmap_and_Dose_DoseRate}a illustrates how such a change would affect the critical doses and dose rates to achieve interspur interaction. Another crucial variable which strongly influences the qualitative results of the model, but whose value remains very uncertain, is the spur decay constant $k_s$ (see Figure \ref{Figures/Heatmap_and_Dose_DoseRate}b). This is discussed in detail in Section \ref{Discussion/Model} alongside other limitations of the model. 

It remains clear that experimental verification of the radiochemical phenomena predicted by this model is necessary to fit the parameters of the model before quantitative predictions can be made. Notwithstanding, the model makes interesting qualitative predictions, provides insight into the irradiation parameters vital to interspur interaction, and produces easily testable hypotheses.

\subsection{Application to FLASH-RT}
\subsubsection*{Parameter Dependencies}
If interspur interactions are responsible for the normal tissue sparing observed in the FLASH effect, one would expect this sparing to have similar parameter dependencies as the onset of significant interspur interactions. Indeed, parameter dependencies predicted by the model, such as a minimum dose and minimum dose rate, are reflected in empirical FLASH studies~\cite{Wilson_2020, Vozenin_2019_SleepingBeauty, Montay-Gruel_2017}. Assuming values of $\Phi=1$ and $k_s=\SI{8e3}{s^{-1}}$, and keeping in mind the aforementioned uncertainties in quantitative prediction by this model, the onset of significant interspur interactions predicted by this model reflects the minimum dose and dose-rate dependencies shown in some empirical FLASH studies of $\approx\SI{10}{Gy}$ and $\approx\SI{40}{Gy/s}$, respectively (see Figure \ref{Figures/Heatmap_and_Dose_DoseRate}a)~\cite{Wilson_2020, Vozenin_2019_SleepingBeauty, Montay-Gruel_2017}.

Most FLASH experiments in the literature, including the one referenced here, use pulsed electron beams, whereby dose rates within each pulse may reach values of $\approx\SI{e6}{Gy/s}$. As is discussed in detail in Section \ref{Discussion/Pulsed}, a pulsed beam greatly increases interspur interaction, further strengthening the probability that interspur interactions play a role in the FLASH effect. In a similar fashion, the very high local dose rates of scanned proton pencil beams should be taken into consideration~\cite{Cunningham_2021}.

Assuming again that interspur interactions are responsible for the FLASH effect, this model makes a few crucial predictions for eliciting a FLASH effect at a fixed total treatment dose.
\begin{itemize}
    \item The total treatment dose must be above a threshold value, regardless of how high the dose rate is.
    \item The dose rate must be increased (i.e. exposure time decreased) beyond a critical value, which depends on the dose (lower doses require higher dose rates), to begin eliciting the FLASH effect.
    \item Increasing the dose rate far beyond that critical value will have diminishing benefits.
    \item The beam LET should be reduced as much as possible; i.e., higher energies at the target should be used for particle beams.
    \item Beams should be pulsed with the lowest number of pulses and lowest pulse widths possible (maximize the dose per pulse).
\end{itemize}

Even if the parameter dependencies of the FLASH effect are indeed well-described by this model, such a correlation cannot rule out the effects of other phenomena besides interspur interaction, such as oxygen depletion or other spur-environment interactions which are also expected to have a dose and dose-rate dependency. In order to distinguish such effects from interspur interactions, careful analyses of the relevant types of radiochemical interactions must be performed. The exact mechanism by which interspur interactions could induce normal tissue sparing is beyond the scope of this work, but may be due to increased radical-radical interactions~\cite{Abolfath_2020, Jansen_2021, Koch_2019}. 

\subsubsection*{Differential Sparing of Normal and Tumor Tissue}
Critical to the FLASH effect is the sparing of normal tissue without a coincidental sparing of tumor tissue; without this differential sparing, the therapeutic benefit of FLASH-RT would be lost. The mechanism by which normal tissue is spared by UHDR irradiation must thus be negated within tumor tissue. With respect to interspur interactions, this may be explained by different effective spur decay constants ($k_s$) in normal and tumor tissue caused by differing oxygen, labile iron, or radical scavenger concentrations, leading to reduced interspur interaction. Alternatively, differing biological responses of normal and tumor tissue to the radiochemical effects of interspur interactions may explain this differential sparing. Determining the exact consequences of these differences is beyond the scope of this work but remains vital to be elucidated by future investigation.

\subsection{Model}
\label{Discussion/Model}
\subsubsection*{Spur Characteristics}
The model presented in this work makes use of multiple assumptions based on the ensemble average of a very large number of spurs, such as the straightness and homogeneity of individual spurs along the beam axis. This assumption is valid in the very densely ionizing radiation limit, but may no longer hold when considering sparsely ionizing radiation; the relevant corrections to the model are described in Section \ref{Discussion/Sparsely}. Another assumption of this model is that the border effects of the radiation field can be neglected, i.e. each spur is assumed to be situated within an infinite, isotropic volume of other spurs. A simple back-of-the-envelope calculation (see Section \ref{Appendix/Border}) indicates that for a square radiation field size of side length $\SI{5}{mm}$, less than $0.03\%$ of spurs would experience such border effects, and can therefore be ignored.

Spurs in this model evolve under a reaction-diffusion equation (Equation \ref{DGL}), which comprises a diffusion term, an intraspur and spur-environment interaction term, and an interspur interaction term. An estimation of quantitative values for the diffusion coefficient $\alpha$ and temporal shift parameter $\tau_0$ is performed in Section \ref{Methods/MC} by analyzing a $\SI{1}{MeV}$ proton spur simulated with the Monte Carlo code TOPAS n-Bio. The simulation yields very good agreement with the spatial variance over time of the model's spur. This particle and energy were chosen due to the straightness of its track, important for the assumption of straight spurs. Although the calculated value of $\alpha$ should be independent of the particle type and energy, since it depends on the simulated diffusion constants of the spur's constituents, the value of $\tau_0$ may be dependent on the radiation quality. For example, spurs of higher-energy charged particles may have a larger initial spatial variance due to physical effects such as the larger ranges of secondary particles, Coulomb explosion, and thermal spike~\cite{Laverne_2000}. Recently, the effects of thermal spike on the shockwave-like outward propagation of chemical species have been described~\cite{Abolfath_2022}, which could induce more interspur interaction and less intraspur interaction compared to when only thermal diffusion is considered. as different particles will exhibit different spatial ionization patterns. However, the effect of $\tau_0$ on $\Phi_{t\gg T}$ in this model, compared to parameters like $k_s$, is minimal; a change over many orders of magnitude ($\SIrange{e-15}{e-6}{s}$) of $\tau_0$ changes $\Phi_{t\gg T}$ by less than one order of magnitude. Therefore, these values for $\alpha$ and $\tau_0$ will be assumed to be valid.

\subsubsection*{Spur Interaction}
In this model, the effects of interspur interactions on the evolution of spurs are ignored under the assumption of weakly-interacting spurs; i.e. assuming interspur interactions do not significantly affect the distributions of species within spurs (see Equation \ref{DGL}). However, once interspur interactions become significant, this assumption is no longer valid. Thus, this simplified model is best used to determine when interspur interactions become significant, but is less accurate at describing the dynamics of these interactions past that point. Quantification of both interspur and intraspur interaction is performed assuming second order reaction kinetics with only one reaction rate constant, $k_r$. Although many relevant reactions do indeed follow second order reaction kinetics, this simplification neglects the different types of interactions that may occur due to the varying spatial distributions of different species within a spur. For instance, the consumption of hydroxyl radicals in intraspur interactions is likely dominated by the self-recombination of hydroxyl radicals due to their high local concentrations within the spur; however, when considering interspur interactions, the reaction of hydroxyl with faster-diffusing species such as $\ce{e_{aq}-}$ or $\ce{H^.}$ is more probable.

\subsubsection*{Intraspur and Spur-environment Interactions}
The intraspur and spur-environment interaction term (see Equation \ref{DGL}) is critical to the outcomes of this model. In this model, this term comprises an exponential decay with decay constant $k_s$. In reality, this decay is likely not exponential as it is a combination of many different processes, e.g. multiple different intraspur reactions, scavenging by different environmental reagents, and production (as opposed to depletion) of different radiochemical species via these reactions. In addition, many of these reactions may be in competition with each other~\cite{Jansen_2021}. The exponential simplification used in this model poses to approximately fit these complex interactions with one useful parameter, the decay constant $k_s$, to qualitatively analyze the effects of varying spur decay due to intraspur or spur-environment interactions. For the same reasons the decay processes are complex and varied, determining an appropriate value for $k_s$ in tissue is non-trivial. In order to get an idea of its approximate value, we examine the decay of a spur simulated in pure water and discuss the effects of scavengers typically found within a cellular environment.

Due to the highly localized concentrations of reactive species within the spur, purely intraspur interaction is initially the most prevalent type of interaction after the creation of a spur. Figure \ref{Figures/Intraspur}b depicts a simulated $\SI{1}{MeV}$ proton spur in $\SI{2}{\mu m}$ of water undergoing solely intraspur interactions (no other spurs and no environmental reagents such as oxygen were included). The spur's original concentration is initially rapidly halved within $\approx \SI{100}{ns}$, after which intraspur interactions barely affect the spur. An exponential decay clearly underestimates the simulated decay rate at very short times while overestimating the decay rate at longer times. Likely, accurately reflecting the spur's decay dynamics in this model at longer time scales is more important than at shorter time scales when interspur interactions are minimal (see Figure \ref{Figures/Electron_Beam_Results}a). To this end, it could be advantageous to represent the effects of intraspur interaction by halving the initial concentration $c_0$ instead of including them in the decay constant $k_s$.

After the initial rapid self-consumption of the spur due to intraspur interactions, spur-environment interactions likely begin to play a larger role. Glutathione is an important hydroxyl radical scavenger, with a scavenging rate of $\approx\SI{e10}{s^{-1}M^{-1}}$~\cite{Mesa_2013, Sueishi_2014}, present in the cell cytosol at concentrations of $\approx\SI{1}{mM}$~\cite{Forman_2009}. Assuming an excess of glutathione, the resulting hydroxyl decay constant would be $\approx\SI{e7}{s^{-1}}$. If we instead consider the scavenging of hydrated electrons by an excess of oxygen (scavenging rate $\approx\SI{e10}{s^{-1}M^{-1}}$~\cite{Buxton_1988}) in a 5\% ($\approx\SI{50}{\mu M}$) oxygen solution to form the superoxide anion $\ce{O_2^.-}$, the decay constant would be $\approx\SI{5e5}{s^{-1}}$. It should also be considered that spur-environment interactions not only consume radicals, but can produce more radicals. For example, stable hydrogen peroxide initially formed by recombination of hydroxyl radicals may later form hydroxyl radicals again via the Fenton reaction, effectively decreasing the decay rate of the spur. Within tissue, many factors such as scavenger concentrations, labile iron concentrations, oxygenation, and pH may affect the spur decay rate. These characteristics may change significantly between tissue types, and thus may differ in different organs and in tumor tissue. Thus, $k_s$ remains a model parameter that may assume a large range of values and can vary to represent different tissue environments. However, as the results of this model are critically dependent on the value of $k_s$ (see Section \ref{Results}), experimental verification of its value is necessary in determining the validity of this model's assumptions.

\subsection{Sparsely Ionizing Radiation}
\label{Discussion/Sparsely}
The assumptions of perfectly straight and homogeneous spurs do not hold for sparsely ionizing (low-LET) radiation. For example, the distribution of ionization events in photon tracks is much more scattered and sparse, and the spatial distinction between individual photon tracks is minimal. In order to address these issues, the model can be adjusted to describe a 3-D distribution of "clusters" of reactive species formed from these ionization events. Their evolution and interaction is handled in a similar way as the original model for densely ionizing radiation. The sparsely ionizing radiation model is detailed in Section \ref{Appendix/Sparsely}.

The results of the sparsely ionizing radiation model have very similar characteristics to that of the densely ionizing radiation model. One important difference is the replacement of the LET ($E_s/z$) with the energy deposited per cluster, $E_c$. Otherwise, there is only a slightly modified dependency on the diffusion constant $\alpha$, the spur decay constant $k_s$, and the minimum spur age $\tau_0$, all of which arise from the change in intraspur interactions from 2-D to 3-D diffusion. The dependencies on time $t$, the exposure time $T$, and the dose $D$ are identical between the two models. Since results are comparable between strictly 2-D, straight spurs of the densely ionizing radiation model and the 3-D, uniformly distributed clusters of the sparsely ionizing radiation model, it can be assumed that realistic spurs of particles, which lay in between the characterizations of these two extreme assumptions, share similar interspur and intraspur interaction behavior. 

\subsection{Pulsed Beam}
\label{Discussion/Pulsed}
Thus far, the construction of this interspur interaction model has only considered a continuous (constant dose rate) beam. However, many FLASH studies take advantage of pulsed beams in order to achieve the ultra-high mean dose rates characteristic of FLASH irradiation. The pulsatile nature of a beam likely critically affects the dynamics of interspur interaction. Assuming a beam of total dose $D$ and total exposure time $T$ pulsed at a frequency $f$ comprising identical pulses of width $w$, the dose per pulse $D_p$ and dose rate within each pulse $\thickdot{D}_p$ can be expressed as
\begin{equation}
    D_p = \frac{D}{f\cdot T} \qquad \text{and} \qquad \thickdot{D}_p = \frac{D}{f\cdot T\cdot w} \ .
    \label{Pulse Equations}
\end{equation}

Each pulse can be considered to be a continuous beam irradiation with total dose $D_p$ and mean dose rate $\thickdot{D}_p$. Thus, by substituting these values for the doses and dose rates depicted in Figure \ref{Figures/Heatmap_and_Dose_DoseRate}, the significance of interspur interaction within each pulse can be analyzed. For a fixed total dose and mean dose rate, decreasing the pulse frequency and decreasing the pulse width results in a higher dose rate per pulse yielding more significant interspur interaction. This approach assumes that interactions between spurs of separate pulses (interpulse) are negligible, and thus represents the lower limit of pulsed-beam interspur interaction. The authors intend to extend this model to include interpulse interaction in order to determine its significance for the irradiation parameters typically used in FLASH studies.

Some of the results shown in Section \ref{Results} reflect the irradiation parameters used in the aforementioned study by Montay Gruel et al.~\cite{Montay-Gruel_2017}, assuming a continuous exposure. However, this beam was actually pulsed, and the onset of the observed FLASH effect occurred when the exposure time was decreased from $\SIrange{0.5}{0.2}{s}$ at a constant pulse frequency of $\SI{100}{Hz}$ and pulse width of $\SI{1.8}{\mu s}$. Thus, the dose per pulse and dose rate per pulse were changed from $\SIrange{0.2}{0.5}{Gy}$ and from $\SIrange{1.1e6}{2.8e6}{Gy/s}$, respectively, to elicit a FLASH effect. Although not shown in Figure \ref{Figures/Heatmap_and_Dose_DoseRate}, a spur decay constant of $k_s=\SI{e5}{s^{-1}}$ would place the isovalue curve of $\Phi_{t\gg T}=1$ at this parameter transition. Considering the pulsatile nature of this beam results in a large increase in the significance of interspur interaction.

\section{Conclusion}
\label{Conclusion}

The objective of this work was to analyze the effects of interspur interactions on the overall radiochemistry at ultra-high dose rates, and examine whether interspur interactions may play a role in the FLASH effect. To this end, an analytical model was developed to describe the spatiotemporal distribution, evolution, and interaction of an ensemble of spurs in a target during and after irradiation. Individual spurs were defined by a 2-D reaction-diffusion equation comprising the radial broadening of the spur due to thermal diffusion and an exponential decay of the spur due to intraspur and spur-environment interactions. The parameters modulating the spur's diffusive broadening were approximated by fitting the theoretical description of the spur to Monte Carlo simulations. The exponential decay was modulated by a model parameter, the spur decay constant $k_s$. The effects of interspur interactions on the evolution of individual spurs were ignored under the assumption of weakly-interacting spurs. Spurs were distributed throughout the target assuming a spatially homogeneous beam with a constant dose rate. Intraspur and interspur interactions were quantified across the ensemble of spurs by their spatio-temporal overlap assuming second-order reaction kinetics.

The ratio of interspur to intraspur interactions, $\Phi$, was then derived as a function of irradiation parameters (e.g. dose, dose rate, LET). Interspur interactions were considered significant to the overall radiochemistry when this ratio reached unity, i.e. when interspur interactions led to as much species conversion as did intraspur interactions ($\Phi=1$). This model demonstrated that $\Phi$ is predominantly dependent on three parameters: the beam fluence, the exposure time, and the spur decay constant. For any set of parameters, a minimum critical dose and dose rate were shown to be necessary to induce significant interspur interactions. This critical dose and dose rate could be reduced by decreasing the LET or spur decay constant. A similar effect could be achieved by lowering the value of $\Phi$ defining significant interspur interactions.

This model demonstrates that significant interspur interaction may occur at the same threshold doses and dose rates characteristic of the observed FLASH effect~\cite{Wilson_2020, Vozenin_2019_SleepingBeauty}, depending on the values of the model parameters $\Phi$ and $k_s$. This model additionally suggests optimal irradiation parameters necessary to induce interspur interactions, and, by extension, potentially the FLASH effect. The differential sparing effect between normal and tumor tissue could be explained by a different effective spur decay constant within the different tissue types. Significant interspur interaction could lead to normal tissue sparing by, for example, increased radical-radical interaction~\cite{Abolfath_2020, Jansen_2021, Koch_2019}. Further analysis of this effect will need to consider interactions between spurs of separate pulses from pulsed beams, as well as the physical spur-broadening effects of high-LET radiation. In addition, the decay rate and dynamics of spurs in this model should be further developed and analyzed for more accurate description. Investigations of this effect will need to discriminate between interspur interactions and other interaction phenomena induced by high dose rates, such as oxygen depletion. Radiochemical analyses of the yields of various radiolytic species in different media as a function of beam parameters are currently underway and will help determine values for the critical parameters of this model.

\newpage
\printbibliography

\newpage
\appendix
\section{Appendix}
\label{Appendix}

\subsection{Spur Reaction Diffusion Equation}
\label{Appendix/PDF}
The general partial differential equation governing the spatio-temporal evolution of spurs undergoing diffusion and linear consumption, ignoring interspur interactions as described in Section \ref{Methods/Definition}, is
\begin{equation}
    \frac{\partial}{\partial t}c(\vec{r},t) = \alpha\nabla^2 c(\vec{r},t) - k_s c(\vec{r},t) \ .
    \label{DGL Appendix}
\end{equation}
Using the Fourier transform
\begin{equation}
    \hat{P}(k_1,...,k_n) = \int_{\textbf{R}^n} \, e^{-2\pi i(k_1 \cdot x_1 + ... + k_n \cdot x_n)}\cdot P(x_1,...,x_n) \ dx...dx_n \ 
\end{equation}
and the identity
\begin{equation}
    \frac{\partial ^n}{\partial x^n}P = (2\pi i k_x)^n \hat{P} \ ,
\end{equation}
Equation \ref{DGL Appendix} can be rewritten as
\begin{equation}
    \frac{\partial}{\partial t}\hat{c} = [-4\alpha\pi^2(k_x^2 + k_y^2) - k_s]\hat{c} \ ,
\end{equation}
thereby losing its spatial dependency. Solving for $\hat{c}$ thus becomes trivial,
\begin{equation}
    \hat{c} = A \cdot e^{[-4\alpha\pi^2(k_x^2+k_y^2) - k_s] \cdot t} \ ,
\end{equation}
and its real-space equivalent can be constructed using the inverse Fourier transform
\begin{equation}
    P(x_1,...,x_n) = \int_{\textbf{R}^n} \, e^{2\pi i(k_1 \cdot x_1 + ... + k_n \cdot x_n)}\cdot \hat{P}(k_1,...,k_n) \ dk...dk_n
\end{equation}
and the initial condition
\begin{equation}
    \int c(\vec{r}, t=0)\, dV=c_0 \ ,
\end{equation}
yielding
\begin{equation}
    c(\vec{r},t) = c_0 \cdot \frac{e^{-\frac{\abs{\vec{r}}^2}{4\alpha t} - k_s t}}{4\pi \alpha t} \ .
    \label{PDF Appendix}
\end{equation}

\subsection{Interspur Interaction Measure}
\label{Appendix/Interspur}

Evaluating Equation~\ref{Summed Average Interaction Rate Density Equation} considering the limit $R\rightarrow\infty$ with the formulations derived in Sections \ref{Methods/Definition}, \ref{Methods/Distribution}, and \ref{Methods/Interaction} yields
\begin{align}
    \sum_i \, \omega_{a,i}(t) &= \lim_{R\rightarrow\infty} N_R(t) \cdot \langle \omega_{1,2}(t) \rangle\\
    & = \lim_{R\rightarrow\infty} N_s(t) \cdot \frac{\pi R^2}{A} \cdot \int \!\!\! \int\!\!\! \int P_t(t_1) \cdot P_t(t_2) \cdot P_s(s) \cdot \omega_{1,2}(t)\ ds \, dt_2 \, dt_1\\
    &  = \lim_{R\rightarrow\infty} \frac{N_s(t)}{A} \cdot \int \!\!\! \int \!\!\! \int_0^R P_t(t_1)\cdot P_t(t_2) \cdot 2 \pi s \cdot \frac{k_r \cdot c_0^2 \cdot  e^{-\frac{s^2}{4\alpha(2t -t_1 -t_2 + 2\tau_0)} - k_s(2t -t_1 -t_2)}}{4\pi \alpha(2t -t_1 -t_2 + 2\tau_0)}\ ds \, dt_2 \, dt_1\\
    & = \frac{N_s(t)}{A} \cdot k_r \cdot c_0^2 \cdot \int \!\!\! \int P_t(t_1)\cdot P_t(t_2) \cdot e^{-k_s(2t -t_1 -t_2)}  \, dt_2 \, dt_1\\
    & = \frac{N_s(t)}{A} \cdot k_r \cdot c_0^2 \cdot \int \!\!\! \int \left(\frac{1}{\min(t,T)}\right)^2 \cdot e^{-k_s(2t -t_1 -t_2)}  \, dt_2 \, dt_1\\
    &= \frac{N_s(t)}{A} \cdot k_r \cdot c_0^2 \cdot \left(\frac{1}{\min(t, T)}\right)^2 \cdot
    \begin{cases} 
        \int_{0}^t \int_{0}^t  e^{-k_s(2t -t_1 -t_2)} \ dt_2 \, dt_1 & \text{for} \quad t\leq T\\[10pt]
        \int_{0}^T \int_{0}^T  e^{-k_s(2t -t_1 -t_2)} \ dt_2 \, dt_1 & \text{for} \quad t> T
    \end{cases}\\
    &= \frac{N_s(t)}{A} \cdot \frac{k_r}{k_s^2} \cdot c_0^2 \cdot \left(\frac{1}{\min(t, T)}\right)^2 \cdot
    \begin{cases}
        (e^{-k_st}-1)^2 & \text{for} \quad t\leq T\\[10pt]
        (e^{k_sT}-1)^2 \cdot e^{-2k_st} & \text{for} \quad t> T
    \end{cases}
\end{align}
Thus, the total interspur conversion of all spurs in the target volume, following Equation \ref{Total Inter-spur Interaction Equation}, is
\begin{equation}
    I_\text{inter}(t) =
    \begin{cases} 
        \lim_{\epsilon\rightarrow 0} \frac{B}{2} \cdot t \cdot \left(\ln(\frac{t}{\epsilon}) - \Gamma(0, 2k_st) + 2\Gamma(0, k_st) + \Gamma(0, 2k_s\epsilon) - 2\Gamma(0, k_s\epsilon)\right) & \text{for} \quad t\leq T\\[10pt]
        I_\text{inter}(T) + \frac{B}{2} \cdot \frac{(e^{k_sT}-1)^2 }{2k_s} \cdot (e^{-2k_s T}-e^{-2k_st}) & \text{for} \quad t> T
    \end{cases} \ ,
    \label{P_inter(t) Appendix}
\end{equation}
where
\begin{equation}
    B = \frac{N_s(t)^2}{A} \cdot \frac{k_r}{k_s^2} \cdot c_0^2 \cdot \left(\frac{1}{\min(t, T)}\right)^2 \ ,
\end{equation}
and $\Gamma(s, x)$ is the incomplete upper gamma function given by
\begin{equation}
    \Gamma(s, x) = \int_x^\infty t^{s-1}e^{-t}\, dt \ .
\end{equation}

\subsection{Consideration of Border Effects}
\label{Appendix/Border}
Assuming the spatial concentration distribution of a spur is described by a bivariate normal distribution (Equation \ref{PDF Arbitrary}), approximately $99.7\%$ of a spur is contained within a radius of $r=3\sigma$ where $\sigma=\sqrt{2\alpha t}$ is the standard deviation of the distribution, i.e. the mean diffusion distance of any one of the spur's species. Assuming $\alpha=\SI{4.3e-9}{m^2 /s}$ and $t=\SI{1}{\mu s}$, this radius is $r\approx \SI{0.3}{\mu m}$. Assuming a spur cannot interact with any other spur that is centered farther than $2r$ away, then, for a homogeneous square radiation field size of side length $\SI{5}{mm}$, at least $99.97\%$ of spurs in the field do not experience any border effects.

\subsection{Model Application to Sparsely Ionizing Radiation}
\label{Appendix/Sparsely}
Here, the calculations performed in Section \ref{Methods}, which applied to densely ionizing radiation (assuming spurs are straight and homogeneous along the beam axis), are adjusted to apply to sparsely ionizing radiation. Under this assumption, each primary particle may create multiple sparsely distributed ionization events, and each ionization event creates a cluster of species. The ionization events are assumed to be uniformly randomly distributed throughout the target volume. This calculation thereby follows the distribution, evolution, and interaction of 3-dimensional clusters as opposed to 2-dimensional spurs.

\subsubsection{Cluster Definition}
The spatio-temporal evolution of a cluster is given by the same reaction diffusion equation as that of a spur (Equation \ref{DGL}) with diffusion coefficient $\alpha$, decay rate $k_s$, and intercluster interaction rate $k_r$:
\begin{equation}
    \frac{\partial}{\partial t}c_i(\vec{r},t) = \alpha\nabla^2 c_i(\vec{r},t) - k_s c_i(\vec{r},t) - k_r \sum_{i \neq j}^{N_s} c_i(\vec{r},t) c_j(\vec{r},t) \ .
\end{equation}
which, solved in 3 spatial dimensions, assuming only weakly-interacting clusters, applying the temporal shift $\tau_0$ to define the cluster's minimum age, and setting $c_0$ to be the initial quantity of species in a cluster, yields the PDF of an arbitrary cluster,
\begin{equation}
    c_i(\vec{r},t) = \frac{e^{-\frac{\abs{\vec{r}-\vec{r_i}}^2}{4\alpha(t-t_i+\tau_0)} -k_s(t-t_i)}}{8(\pi \alpha (t-t_i+\tau_0))^{3/2}} \ .
\end{equation}

\subsubsection{Cluster Distribution}
Assuming a beam of constant dose rate over an exposure time $T$, the PDF of the creation time $t_i$ of the $i$th cluster at time $t$ is
\begin{equation}
    P_t(t_i) =
    \begin{cases} 
      \frac{1}{\min(t, T)} \quad & \text{for} \quad 0\leq t_i\leq \min(t,T)\\
      0 \quad & \text{elsewhere}
   \end{cases}
   \ .
   \label{P_t(t_i) Cluster Equation}
\end{equation}
The expected total number of clusters in the target over time $N_c(t)$ can be expressed with the total energy deposited in the target $E$ and the average energy deposited per cluster $E_c$ for a target of density $\rho$ and volume $V$: 
\begin{align}
    N_c(t) &= \frac{E}{E_c} \cdot \frac{\min(t, T)}{T}\\
    &= \frac{D \cdot V \cdot \rho}{E_c} \cdot \frac{\min(t, T)}{T} \ .
\end{align}
$E_c$ will depend on the radiation quality.

Assuming a spatially homogeneous beam, clusters arrive uniformly at random throughout the target volume, so the expected number of clusters within any spherical volume of radius $R$ can be expressed as a fraction of the total number of clusters on the target:
\begin{equation}
    N_R(t) = \frac{4\pi R^3}{3V} \cdot  N_c(t) \\
    \label{N_R(t) Cluster Equation}
\end{equation}
Similarly, the PDF of the displacement $s$ between one arbitrary cluster and another cluster within a radius $R$ is given by the ratio between the volume element $4\pi r^2$ and the total volume $\frac{4}{3}\pi R^3$:
\begin{equation}
    P_s(s) = 3\frac{r^2}{R^3} \ , \quad s \leq R \ .
    \label{P(mu) Cluster Equation}
\end{equation}

\subsubsection{Cluster Interaction Measure}
\label{AppendixB/Sparsely/Interaction Measures}
The interaction rate $\omega$ between two individual clusters is:
\begin{align}
    \omega_{1,2}(t) &= \int k_r \cdot c_1(\vec{r},t) \cdot c_2(\vec{r},t) \, dV\\
    &= k_r\cdot c_0^2 \cdot \frac{e^{-\frac{\abs{\vec{r_1}-\vec{r_2}}^2}{4\alpha(2t - t_1 - t_2 + 2\tau_0)} - k_s(2t - t_1 - t_2)}}{8( \pi \alpha (2t - t_1 - t_2 + 2\tau_0))^{3/2}}  \ .
    \label{Interaction Rate Density Cluster Equation}
\end{align}

\subsubsection{Intercluster Interactions}
The total intercluster interaction rate of an arbitrary cluster $a$ with all neighboring clusters is therefore
\begin{align}
    \sum_i \omega_{a,i}(t) &= \lim_{R\rightarrow\infty} N_R(t) \cdot \langle \omega_{1,2}(t) \rangle\\
    & = \lim_{R\rightarrow\infty} N_c(t) \cdot \frac{4 \pi R^3}{3V} \cdot \int \!\!\! \int\!\!\! \int P_t(t_1) \cdot P_t(t_2) \cdot P_s(s) \cdot \omega_{1,2}(t)\ ds \, dt_2 \, dt_1\\
    &= \frac{N_c(t)}{V} \cdot k_r \cdot c_0^2 \cdot \left(\frac{1}{\min(t, T)}\right)^2 \cdot
    \begin{cases} 
        \int_{0}^t \int_{0}^t  e^{-k_s(2t-t_1-t_2)} \ dt_2 \, dt_1 & \text{for} \quad t\leq T\\[10pt]
        \int_{t-T}^t \int_{t-T}^t  e^{-k_s(2t-t_1-t_2)} \ dt_2 \, dt_1 & \text{for} \quad t> T
    \end{cases}\\
    &= \frac{N_c(t)}{V} \cdot \frac{k_r}{k_s^2} \cdot c_0^2 \cdot \left(\frac{1}{\min(t, T)}\right)^2 \cdot
    \begin{cases} 
        (e^{-k_st}-1)^2 & \text{for} \quad t\leq T\\[10pt]
        (e^{k_sT}-1)^2 \cdot e^{-2k_st} & \text{for} \quad t> T
    \end{cases}
\end{align}

Thus, the total intercluster conversion is 
\begin{align}
    I_\text{inter}(t) &= \frac{N_c(t)}{2} \cdot \int_{0}^t \lim_{R\rightarrow\infty} N_R(t') \cdot \langle \omega_{1,2}(t') \rangle \, dt' \\
    &=
    \begin{cases} 
        \lim_{\epsilon\rightarrow 0} \frac{B}{2} \cdot t \cdot \left(\ln(\frac{t}{\epsilon}) - \Gamma(0, 2k_st) + 2\Gamma(0, k_st) + \Gamma(0, 2k_s\epsilon) - 2\Gamma(0, k_s\epsilon)\right) & \text{for} \quad t\leq T\\[10pt]
        I_\text{inter}(T) + \frac{B}{2} \cdot \frac{(e^{k_sT}-1)^2 }{2k_s} \cdot (e^{-2k_s T}-e^{-2k_st}) & \text{for} \quad t> T
    \end{cases} \ ,
    \label{P_inter(t) Appendix Cluster}
\end{align}
where
\begin{equation}
    B = \frac{N_c(t)^2}{A} \cdot \frac{k_r}{k_s^2} \cdot c_0^2 \cdot \left(\frac{1}{\min(t, T)}\right)^2 \ .
\end{equation}

\subsubsection{Intracluster Interactions}
The total intracluster conversion in a continuous beam is approximated using the same method as described in Section \ref{Methods/Instraspur},
\begin{align}
    I_\text{intra}(t) &= \sum_i^{N_c(t)} \int_{0}^{t}  \omega_{i,i}(t) \, dt'\\
    &\approx N_c(t) \cdot \int_0^\infty \omega_{i,i}(t) dt\\
    &= N_c(t) \cdot \int_0^\infty k_r\cdot c_0^2 \cdot \frac{e^{-2k_s t}}{8(2\pi \alpha(t + \tau_0))^{3/2}} \, dt  \\
    &= N_c(t) \cdot k_r \cdot c_0^2 \cdot \frac{\sqrt{k_s}}{16(\pi \alpha)^{3/2}} \cdot e^{2k_s\tau_0} \cdot \Gamma\left(-\frac{1}{2},\,2k_s\tau_0\right)
\end{align}

\subsubsection{Comparison to Densely Ionizing Radiation}
In order to compare the sparsely ionizing radiation application of this model with clusters to the densely ionizing radiation application with spurs, the subscripts "clusters" and "spurs" will be used to denote the relevant quantities, which differ as follows:
\begin{align}
    I_\text{inter, clusters}(t) &= I_\text{inter, spurs}(t) \cdot \frac{L^2}{E_c^2} \cdot \frac{V}{A} \\
    I_\text{intra, clusters}(t) &= I_\text{intra, spurs}(t) \cdot \frac{L}{E_c} \cdot \frac{V}{A} \cdot \sqrt{\frac{k_s}{4\pi\alpha}} \cdot \frac{\Gamma(-\frac{1}{2},\, 2k_s \tau_0)}{\Gamma(0,\, 2k_s\tau_0)} \\
    \Rightarrow \Phi_{\text{clusters}}(t) &= \Phi_{\text{spurs}}(t) \cdot \frac{L}{E_c} \cdot \sqrt{\frac{4\pi\alpha}{k_s}} \cdot \frac{\Gamma(0,\, 2k_s\tau_0)}{\Gamma(-\frac{1}{2},\, 2k_s\tau_0)} \ .
\end{align}

\end{document}